\documentclass{aastex61}

\usepackage{amssymb}
\usepackage{amsmath}
\usepackage{graphicx}
\newcommand{\Lagr}{\mathcal{L}}

\newcommand{\mdp}{\textrm{MDP}}

\shorttitle{}
\shortauthors{Lowell et al.}

\begin{document}

\title{Polarimetric Analysis of the Long Duration Gamma Ray Burst GRB 160530A With the Balloon Borne Compton Spectrometer and Imager}


\author{A.W. Lowell, S.E Boggs, C.L. Chiu, C.A. Kierans, C. Sleator, J.A. Tomsick, A.C. Zoglauer}
\affil{Space Sciences Laboratory, University of California, Berkeley,  USA}
\author{H.-K. Chang, C.-H. Tseng, C.-Y. Yang}
\affil{Institute of Astronomy, National Tsing Hua University, Taiwan}
\author{P. Jean, P. von Ballmoos}
\affil{IRAP Toulouse, France}
\author{C.-H. Lin}
\affil{Institute of Physics, Academia Sinica, Taiwan}
\author{M. Amman}
\affil{Lawrence Berkeley National Laboratory, USA}

\begin{abstract}

A long duration gamma-ray burst, GRB 160530A, was detected by the Compton Spectrometer and Imager (COSI) during the 2016 COSI Super Pressure Balloon campaign.  As a Compton telescope, COSI is inherently sensitive to the polarization of gamma-ray sources in the energy range 0.2-5.0 MeV.  We measured the polarization of GRB 160530A using 1) a standard method (SM) based on fitting the distribution of azimuthal scattering angles with a modulation curve, and 2) an unbinned, maximum likelihood method (MLM).  In both cases, the measured polarization level was below the $99\%$ confidence minimum detectable polarization levels of $72.3 \pm 0.8$\% (SM) and $57.5 \pm 0.8\%$ (MLM).  Therefore, COSI did not detect polarized gamma-ray emission from this burst.  Our most constraining 90\% confidence upper limit on the polarization level was 46\% (MLM).  

\end{abstract}
\keywords{balloons --- instrumentation: polarimeters --- gamma rays: general --- gamma-ray burst: individual (GRB 160530A) --- polarization --- techniques: polarimetric}

\section{Introduction} \label{sec:intro}

Decades of broad band observations of gamma-ray bursts (GRBs) have shed light on the nature of GRBs and their progenitors.  Despite this, there are still significant gaps in our understanding of the GRB prompt gamma-ray emission.  In particular, the emission mechanism, emission geometry, and magnetic field structure of the inner GRB jet are currently not well understood.  A major challenge that any model for the GRB prompt emission faces is reconciling the spectral and temporal diversity that is seen across the GRB population.  It is believed that polarization measurements, taken together with spectra and light curves, have the potential to significantly further our understanding of the GRB prompt emission.

Various scenarios have been proposed which can yield a net linear polarization of the prompt emission, as measured by a distant observer \citep{toma2009}.  Synchrotron emission is often invoked as the emission mechanism, which can produce high levels of polarization when an ordered, large-scale magnetic field exists in the jet and the viewing angle happens to fall within the jet cone \citep{lyutikov2003,granot2003}.  If the magnetic field is randomly oriented over small length scales throughout the jet, then a high level of polarization can only be seen when the viewing angle aligns with the edge of the jet, due to a loss of symmetry \citep{gruzinov1999,waxman2003}.  Likewise, inverse Compton emission can yield high levels of linear polarization, but only when the viewing angle is near the jet edge \citep{shaviv1995,lazzati2004}.  In general, each scenario can yield a range of polarization levels which depends on the observer viewing angle, as well as jet opening angle, the Lorentz factor of the burst, spectral parameters, etc.  Therefore, in order to discriminate between models, a statistical analysis of a large sample of GRB polarization measurements is required \citep{toma2009}.

To date, only a limited number of GRB prompt emission polarization measurements exist, with detection significances ranging from $\sim 1.5 - 4 \sigma$ \citep{mcconnell2017}.  Most of these measurements were carried out with instruments that were not optimized or calibrated to perform gamma-ray polarimetry.  One exception is IKAROS/GAP \citep{yonetoku2011}, a dedicated GRB polarimeter aboard the JAXA solar sail IKAROS which is responsible for the most significant GRB polarization measurements to date \citep{yonetoku2012}.  The POLAR instrument aboard the Chinese space station Tiangong 2 is another dedicated GRB polarimeter which was only recently deployed \citep{sun2016,orsi2014,produit2005}.  POLAR has successfully detected several GRBs in its few months of service \citep{polargcn1,polargcn2}, although no polarization analyses have yet been published.  Most recently, the Astrosat/CZTI team \citep{czti} has reported GRB polarization measurements for eleven bright GRBs with detection significances ranging from $2.5\sigma$ to less than $4\sigma$ \citep{astrosatgrb}.

The Compton Spectrometer and Imager (COSI) is a 0.2-5.0 MeV gamma-ray Compton imager, spectrometer, and polarimeter developed under the NASA APRA program to fly on NASA's newly developed 18 million cubic feet Super Pressure Balloon (SPB) \citep{chiu2015,kierans2017}.  COSI combines a wide field of view, 3D positioning, and high spectral resolution in order to address a number of science targets, including GRB polarization.  Here we report the COSI detection of the bright, long-duration GRB 160530A and the associated polarimetric analysis for this GRB.

\section{The Compton Spectrometer and Imager} \label{sec:cosi}

The COSI detector system is comprised of a $2 \times 2 \times 3$ array of high-purity germanium double-sided strip detectors (GeDs) \citep{amman2007}  surrounded by a scintillating cesium iodide (CsI) anti-coincidence shield (ACS).  Each GeD measures 8~cm on a side and 1.5 cm thick, with the electrodes on each side segmented into 37 strips.  3D positioning in the GeDs is achieved by identifying the $x$ and $y$ strips that triggered on each side of the detector, and then converting the time difference between the $x$ and $y$ signals to infer the depth, or $z$ coordinate \citep{lowell2016}.  The position resolution in the $x$ and $y$ directions is equal to the strip pitch, which is 2 mm.  Along the $z$ direction, the position resolution is pixel dependent, as it depends primarily on electronic noise, but on average is 0.2 mm rms.  High spectral resolution is achieved by virtue of using germanium as the detector material, with an average energy resolution of 2.7 keV at 662~keV per channel.  

An ideal photon event consists of at least one Compton scatter and a subsequent photoelectric absorption, all taking place in the GeDs.  The sequence of events takes place over too small a volume to exploit time of flight for event sequence determination, so other methods are used to predict the true sequencing \citep{zoglauerthesis}.  A sequenced Compton event uniquely specifies a Compton cone whose surface represents the possible points of origin of the photon.  After event reconstruction, Compton events are used in high level analysis such as List Mode Maximum Likelihood Expectation Maximization (LM-MLEM) imaging \citep{wilderman1998} and polarization analysis.  

The predecessor to COSI was named the Nuclear Compton Telescope (NCT).  NCT was flown from Ft. Sumner, New Mexico in 2005 \citep{boggs2006} and 2009 \citep{bandstra2009spring}. The 2009 flight yielded an image of the Crab nebula, which was the first image of an astrophysical source using a compact Compton telescope \citep{bandstracrab}.  Between 2010 and 2014, our collaboration developed COSI which is upgraded in several important ways: 1) A mechanical cryocooler replaced the liquid nitrogen cooling system, thus lifting the constraint of a cryogen-limited flight duration, 2) two more GeDs were added and a new GeD arrangement of $2 \times 2 \times 3$ was chosen so as to optimize the polarization response, 3) the BGO ACS was replaced with a CsI ACS, 4) a lighter weight gondola was constructed to match the lift specification of the SPB, and 5) the previous pointing system was retired in favor of a passive zenith pointing strategy.  In its current configuration, COSI's field of view spans $\sim 25\%$ of the sky\footnote{The field of view is limited by the CsI ACS, which is required for background suppression.  Without the ACS, the field of view would be $4\pi$ steradians.}.  In December of 2014, COSI was launched on a SPB from McMurdo Station, Antarctica.  The instrument performed well and successfully measured the background environment for 36 hours, at which point a leak in the balloon was discovered, effectively ending the flight prematurely.  The payload was successfully recovered, rebuilt, and launched again from Wanaka, New Zealand on 2016 May 16 \citep{kierans2017}.  For 46 days, COSI floated at a nominal altitude of $\sim 33$ km, continuously observing and telemetering all Compton events in real-time over the Iridium satellite network.  Finally, on 2016 July 5, COSI made a relatively gentle landing in the Atacama desert of Peru.  Recovery efforts were successful, including the recovery of the solid state drives containing the full raw data set.

\subsection{Data Pipeline and Simulations}

As photons interact in the COSI detector system, they deposit varying amounts of energy under various strips of the GeDs.  The readout electronics record the $x$ and $y$ strip IDs, pulse heights (energy), and timing (interaction depth) of triggered events and forward these data to the flight computer for storage, real-time analysis, and telemetry.  Simulating the measurement process is crucially important for benchmarking instrument performance, as well as determining instrumental responses.  Thus, we used GEANT4 \citep{geant4} via the software library MEGAlib \citep{zogmegalib} to perform Monte Carlo (MC) simulations of particles propagating through a realistic mass model of the COSI instrument.  \citet{sleator2017} contains a complete description of the COSI data pipeline.  The pipeline is largely the same for measured data and simulated data, with the exception that the simulated data are processed by the ``detector effects engine'' (DEE) \citep{sleator2017} before passing through the calibration modules.  The goal of the DEE is to add as much realism (noise, thresholds, electronics limitations, etc.) as possible to the idealized simulation events.  This way, measurements may be benchmarked against simulations.  At the end of the pipeline, the events -- measured or simulated -- have been reconstructed and are ready for use in high level analysis.  

Data analysis proceeds by imposing event selections which act to reduce background, increase signal, and optimize imaging, spectral, or polarization performance.  The most relevant event selections for polarization analysis are as follows: 
\begin{enumerate}
\item Total photon energy.  All energy deposits are summed to get the total photon energy.  It is possible for a photon to escape the detector system without depositing all of its energy, but these events can be identified with reasonable certainty and are rejected during event reconstruction.
\item First Compton scattering angle.  This is the Compton scattering angle of the first interaction, and is determined by evaluating the classic Compton scattering formula with the total photon energy and the photon energy after the first Compton scatter.
\item Angular resolution measure (ARM).  This quantity is defined as the smallest angular distance between an event's Compton cone and a position in the sky. It is essentially an imaging cut and is applied in order to extract events whose Compton cones are consistent with a location in the sky.  ARM values can be positive or negative, depending on whether the source location falls outside (positive) or inside (negative) the Compton cone.
\item Distance between first two interactions.  Events with large distances are less affected by the GeD position uncertainty and generally yield lower (better) ARM values.
\item Distance between any two interactions.  Events with interactions that are too close are harder to reconstruct properly, so this cut acts to improve the reconstruction efficiency. 
\item Number of interaction sites.  This parameter impacts the reconstruction efficiency, since events with more interaction sites can generally be reconstructed more accurately.  Events with two interaction sites are the most common, followed by three sites, four sites, etc.
\end{enumerate}
Any other event parameter that is not present in this list was not used for event selection.

\section{GRB 160530A} \label{sec:grb}

On May 30, 2016, 14 days into the flight, COSI detected the bright, long duration gamma-ray burst GRB 160530A \citep{cosigcn}.  At the time, COSI's geographic coordinates were lat = $-56.79^{\circ}$, lon = $82.31^{\circ}$, which resulted in a high background level due to the proximity of the South Magnetic Pole.  Additionally, a relativistic electron precipitation event (REP) \citep{parksrep} coincided with the GRB observation, resulting in low-frequency variations in the background count rate. 

\begin{figure}[t]
\includegraphics[width=\textwidth]{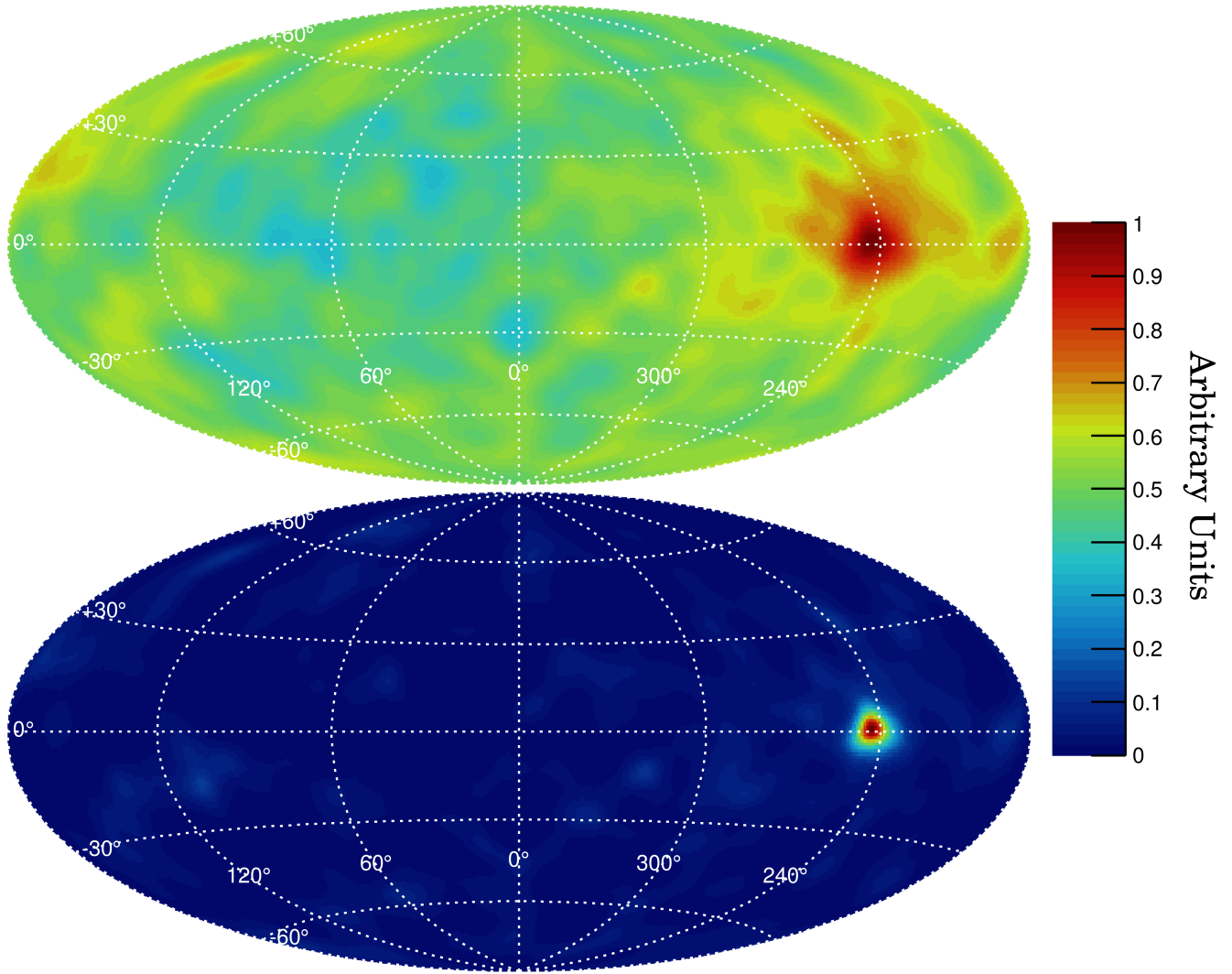}
\caption{Two COSI images of GRB 160530A in Galactic coordinates using zero iterations (top) and ten iterations (bottom) of the LM-MLEM image deconvolution algorithm.  Zero  iterations corresponds to a simple back-projection of the Compton cones.  The event selections used to make this image were different than those used for the polarization analysis, as the goal here is imaging performance.  The color scale intensity is in arbitrary units.} 
\label{fig:mlem}
\end{figure}

Two COSI images of GRB 160530A are shown in Figure \ref{fig:mlem}, where the top image corresponds to a simple back-projection of the Compton cones, and the bottom image corresponds to the same image after 10 iterations of the LM-MLEM algorithm.  The images were made using 834 events with event selections that were optimized for imaging performance. The peak value in the image occurs at $l = 243.4^{\circ}$, $b = 0.4^{\circ}$ ($1.5^{\circ}$ error circle, 90\% confidence), which is the best known localization for this GRB\footnote{This position is different than the position originally reported in \cite{cosigcn} of $l = 243.9^{\circ}$, $b = 2.1^{\circ}$ due to changes and improvements in the event reconstruction algorithm.}. In the local COSI coordinate frame, the GRB was $43.5^{\circ}$  off-axis with an azimuth of $-66.1^{\circ}$, although a gradual azimuthal motion due to an anomaly in the gondola rotator caused the GRB position to shift in the local frame by approximately $1.7^{\circ}$ over the duration of the GRB.  The average position in the local frame was used when performing simulations of GRB 160530A.

GRB 160530A was detected by Konus-Wind and INTEGRAL/ACS \citep{triangulationgcn}, as well as Astrosat/CZTI and the CsI shield system for Astrosat/CZTI (private communication).  Konus and INTEGRAL/ACS are constituents of the Inter-Planetary Network (IPN), a network of gamma-ray detectors scattered around the solar system which use time delays for source triangulation.  IPN carried out a partial triangulation of GRB 160530A using the time delay between the Konus and INTEGRAL/ACS lightcurves.  With only two IPN instruments detecting this GRB with a large baseline, IPN was only able to localize the source position to an annulus in the sky.  The $1.5^{\circ}$ radius COSI error circle (90\% confidence) was found to overlap a region of the IPN annulus, as can be seen in \citet{triangulationgcn}.  This prompted the COSI team to trigger a Swift target of opportunity observation to tile the overlapping region in search of the GRB afterglow between 19 ks and 158 ks post-burst \citep{swift1gcn,swift2gcn,swift3gcn}.  Unfortunately, Swift/XRT did not detect any fading X-ray sources, and therefore no afterglow was detected.  This could be due to several reasons, the most likely being that the reported COSI error circle in \citet{cosigcn} was not accurate at the time, and that improvements in the data pipeline since then would give a better position.  We note that our updated position of $l = 243.4^{\circ}$, $b = 0.4^{\circ}$ still overlaps with the IPN annulus, but that the image in \citet{triangulationgcn} is no longer up to date. 

\begin{figure}[t]
\includegraphics[width=\textwidth]{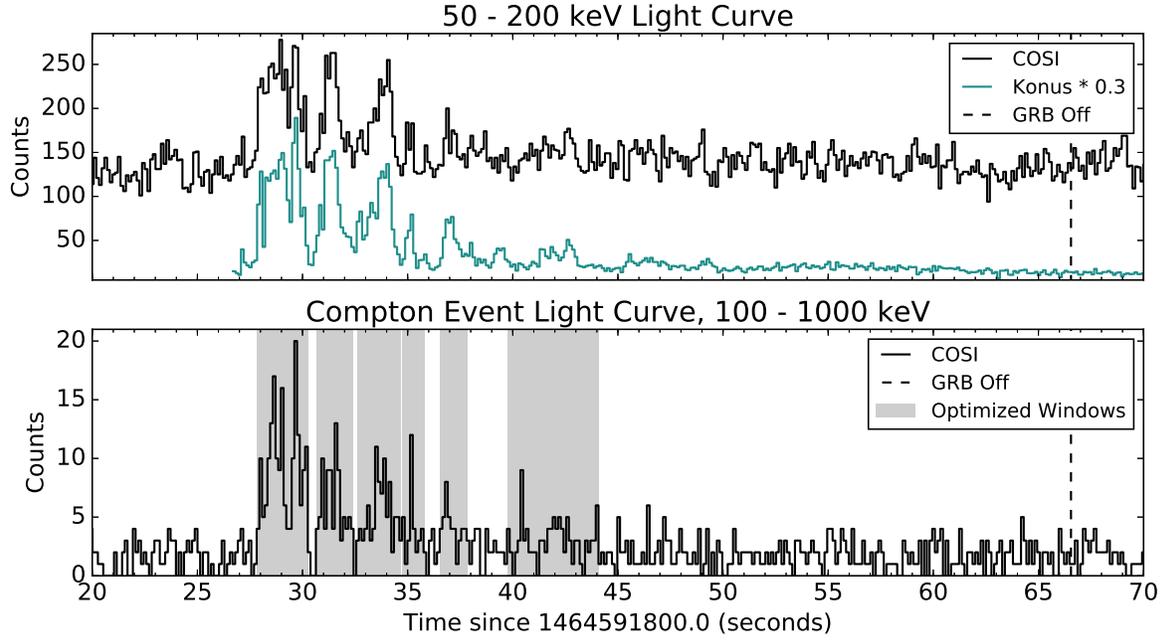}
\caption{(top) COSI and Konus light curves plotted for photon energies between 50 and 200 keV.  The Konus light curve has been shifted by +0.651 seconds to account for the COSI-Konus delay time.  In this energy band, only 2\% of events are Compton events; the rest are single-site events. (bottom) The COSI Compton event light curve, using the event selections shown in Table \ref{table:mlm}.}
\label{fig:lc}
\end{figure}

Figure \ref{fig:lc} shows the COSI and Konus 50 - 200 keV light curves (top), as well as the COSI 100 - 1000 keV Compton event light curve (bottom).  The light curve is characterized by two phases: a decaying pulse train lasting  $\sim 16$ seconds, and an ensuing period of low-level activity lasting $\sim 21$ seconds.  The pulse train is clearly visible in all three light curves, while the low-level activity is only clearly detected in the Konus data.  As a result, we excluded the low-level activity phase from our analysis and only used the six pulses that are clearly seen in the COSI Compton event light curve.  We performed a time domain optimization of the Compton event light curve in order to extract signal from the pulses and reject background from between the pulses.  The procedure for this was to define six time windows bracketing the main pulses, and then use the Differential Evolution algorithm \citep{Storn1997} to find the window start and end times that optimize the significance $S/\sqrt{S + B}$, where $B$ is the expected number of background counts and $S$ is the estimated number of source counts.  The quantity $B$ is estimated by calculating the average background count rate from two 200 second time periods bracketing the GRB, and multiplying the background rate by the total time spanned by the windows.  Subtracting $B$ from the total number of counts within the time windows yields $S$.  The optimized time windows are shown as the grey regions in the bottom panel of Figure \ref{fig:lc}, and printed in Table \ref{table:times}.  The total time spanned by the optimized windows is 12.945 seconds.

\begin{table}
\begin{center}
\begin{tabular}{ |c|c|c| }
\hline
 Pulse & Start Time & Stop Time \\ \hline
 1 & 27.845 & 30.286 \\ \hline
 2 & 30.650 & 32.374 \\ \hline
 3 & 32.607 & 34.653 \\ \hline
 4 & 34.738 & 35.825 \\ \hline
 5 & 36.518 & 37.826 \\ \hline
 6 & 39.731 & 44.070 \\ \hline
\end{tabular}
\end{center}
\caption{Optimized time windows in units of seconds.  Times are with respect to UTC = 1464591800 seconds (May 30th, 2016 07:03:20 UTC). The total time spanned by the windows is 12.945 seconds.}
\label{table:times}
\end{table}

The time delay between the COSI light curve and the Astrosat/CZTI shield light curve was used to constrain COSI's absolute timing accuracy.  COSI events are time tagged using a 10 MHz oscillator and then converted to an absolute time using the pulse-per-second signal from a GPS receiver.  In theory, the absolute timing accuracy of the system is $<1 \mu$s.  However, calibrating the timing accuracy to this precision is challenging, and systematic effects will worsen this figure.  The best fit time delay between COSI and the Astrosat/CZTI shield was $1 \pm 5$ ms (K. Hurley, private communication), with COSI detecting the GRB earlier.  Using the position of the COSI and Astrosat instruments along with the best known COSI position for GRB 1605030A, the time delay should be 5 ms, with Astrosat detecting the burst first.  Given that the absolute timing accuracy of Astrosat is known to be $2 \pm 0.3$ ms (D. Bhattacharya, private communication), the difference between the observed and predicted time is then $6 \pm 5$~ms, where the time uncertainties have been added in quadrature.  Therefore, We conclude that that the absolute timing accuracy of COSI is good to 11 ms at a confidence level of 1$\sigma$.

The Konus team performed a spectral analysis of GRB 160530A on 38.9 seconds of data between 20 keV and 5 MeV \citep{spectrumgcn}.  Fitting with a Band model yielded $\alpha = -0.93 \pm 0.03$, $E_{p} = 638^{+36}_{-33}$ keV, and an upper limit on $\beta$, $\beta < -3.5$ (errors are 90 \% confidence).  The 20 keV to 10 MeV fluence was reported to be $1.30 \pm 0.04 \times 10^{-4}$~erg~cm$^{-2}$.  For our MC simulations of GRB 160530A, we used the best fit Konus Band model \citep{band} with $\beta = -3.5$, along with a multiplicative absorption model to account for extinction in the atmosphere.  The effective column density was determined by integrating the NRLMSISE00 \citep{nrlmsise} model along the line of sight towards the GRB through a spherical atmosphere out to an altitude of 100 km.  The result was an effective column density of 9.2 g cm$^{-2}$.  A spectral analysis of GRB 160530A using the COSI data set will be presented elsewhere (Sleator et al., in preparation).

\section{Polarimetric Analysis} \label{sec:analysis}

The method for measuring the polarization level and angle of a beam in the Compton regime (100 keV to 10 MeV) -- where the GRB prompt emission energy output peaks -- is referred to as Compton polarimetry.  When a photon Compton scatters, the scattered photon is more likely to scatter in a direction that is perpendicular to the incident photon's electric field vector \citep{lei1997}.  In other words, if $\eta$ is the azimuthal scattering angle, where $\eta = 0^{\circ}, 180^{\circ}$ corresponds to scattering along the electric field vector, then the photon preferentially scatters such that $\eta = -90^{\circ}$ or $+90^{\circ}$.  Therefore, a Compton polarimeter must be able to localize the first and second interaction locations so that the azimuthal scattering angle can be measured geometrically.  For a polarized beam, some fraction of the photons will have their electric field vectors aligned with a specific orientation, and so the statistical distribution of the azimuthal scattering angle is used to infer the polarization level and angle of the beam.

Two different methods were used to determine the polarization level and angle of GRB 160530A: a standard method (SM), where the azimuthal scattering angles are histogrammed and then fit with a ``modulation'' curve, and a maximum likelihood method (MLM) which does not bin the data and uses more information per photon.  A detailed description of our implementation of the MLM can be found in the accompanying paper ``Maximum Likelihood Compton Polarimetry with the Compton Spectrometer and Imager,'' hereafter referred to as P2 (Lowell et al. 2017, submitted).

\newpage
\subsection{Standard Method} \label{sec:sm}

\begin{table}
\begin{center}
\begin{tabular}{ |c|c| }
\multicolumn{2}{c}{Event Selections} \\ \hline
Total photon energy & $111 - 847$ keV \\ \hline
First Compton scattering angle & $55.4^{\circ} - 145.2^{\circ}$ \\ \hline
Angular resolution measure & $|\textrm{ARM}| < 12.5^{\circ}$\\ \hline
Smallest distance between first two interactions & 0.5 cm \\ \hline
Smallest distance between any two interactions & 0.3 cm \\ \hline
Number of interaction sites & 2 - 7 \\ \hline
\multicolumn{2}{c}{} \\
\multicolumn{2}{c}{Statistics} \\ \hline
Total counts $T$ & 445 \\ \hline
Background counts $B$ & 123 \\ \hline
MDP & $72.3 \pm 0.8\%$ \\ \hline
Modulation factor $\mu_{100}$ & $0.484 \pm 0.002$ \\ \hline
\multicolumn{2}{c}{} \\
\multicolumn{2}{c}{Fit Results} \\ \hline
Amplitude (counts s$^{-1}$) & $0.12  \pm  0.08$ \\ \hline
Offset (counts s$^{-1}$) & $0.76 \pm 0.05$ \\ \hline
Fitted polarization angle $\hat{\eta_0}$ & $117^{\circ +19^{\circ}}_{-20^{\circ}}$\\ \hline
Fitted modulation $\hat{\mu}$ & $0.16^{+0.16}_{-0.15}$ \\ \hline
Polarization level $\Pi = \hat{\mu} / \mu_{100}$ & $33^{+33}_{-31}\%$ \\ \hline
$\Pi$ 90\% upper limit & 79\% \\ \hline
$\chi^{2}_{\rm red}$ (dof = 27) & 0.98 \\ \hline
\end{tabular}
\end{center}
\caption{SM event selections, statistics, and fit results.  The number of bins in the azimuthal scattering angle distribution was 30 ($12^{\circ}$ per bin).}
\label{table:sm}
\end{table}

The polarization level and angle are determined in the SM by fitting the following function to the azimuthal scattering angle distribution:

\begin{equation}
F - A\cos(2(\eta - \eta_0)),
\label{eq:smfitfunc}
\end{equation}
where $F$ is the offset, $A$ is the amplitude, and $\eta_0$ is the polarization angle.  During the $\chi^2$ minimization, $A$, $F$, and $\eta_0$ are left free.  The measured modulation $\hat{\mu}$ is defined as $\hat{\mu} = A/F$.  To convert the measured modulation to a polarization level, $\hat{\mu}$ must be divided by the modulation factor $\mu_{100}$, which is the modulation in the case that the source is 100\% polarized.

The first step in the SM is to determine the event selections that optimize the statistical minimum detectable polarization \citep{weisskopf,strohmayer}:
\begin{equation}
\textrm{MDP} = \frac{4.29}{\mu_{100} r_s} \cdot \sqrt{\frac{r_s + r_b}{t}},
\label{equation:mdp}
\end{equation}
where $r_s$ is the average source count rate, $r_b$ is the average background count rate, and $t$ is the observation time.  The factor of 4.29 in Equation \ref{equation:mdp} corresponds to a confidence level of 99\%.  During the optimization, the observation time $t$ was fixed at 12.945 seconds (from the time domain optimization), while $r_s$, $r_b$, and $\mu_{100}$ were computed for various energy, Compton scattering angle, and ARM selection ranges.  For each ensemble of event selections, $r_s$ and $r_b$ were computed using the real data, while $\mu_{100}$ was computed using a polarized simulation of GRB 160530A with an arbitrary polarization angle.  We employed the Differential Evolution optimization algorithm to find the energy range, Compton scattering angle range, and ARM cut that optimize Equation \ref{equation:mdp}.  It is reasonable to expect the range on the Compton scattering angle to close in towards $\sim 90^{\circ}$, as this is where the ideal modulation peaks \citep{lei1997}, giving a larger value for $\mu_{100}$.  However, a trade-off exists between the magnitude of $\mu_{100}$ and the source count rate; the optimized event selections should give as large a value for $\mu_{100}$ as possible while still accepting as many source counts as possible. The other event selections -- distance between interactions, and number of interaction sites -- were fixed during the optimization.  The first distance cut was set to 0.5 cm so as to avoid artifacts that appear in the azimuthal scattering angle distributions when the interactions are too close.  These artifacts are a result of the finite position resolution of the detectors.  The cut on the minimum distance between any two interactions was set at a more relaxed 0.3 cm, which has the effect of rejecting many 3+ site events where interactions occur on neighboring strips (strip pitch = 0.2 cm).  This has no impact on the azimuthal scattering angle measurement, since the azimuthal angle is only measured using the first two interactions which are subject to the more restrictive 0.5 cm cut.  Finally, events with two or more interaction sites were used in order to get as many counts as possible, although no events were detected with more than five interaction sites.  Table \ref{table:sm} shows the event selections used for the SM. 

After applying these selections, $T = 445$ total counts remained for the analysis, approximately $B = 123$ of which were background.  The number of background counts $B$ was estimated by fitting a third order polynomial to the light curve 200 seconds before and after the GRB, but with the time bins corresponding to the GRB excluded from the fit.  Integrating the polynomial over the optimized time windows resulted in $B = 123$.  Using Equation \ref{equation:mdp} and assuming no systematic error, the MDP was $58 \pm 2\%$.  In order to determine the MDP including systematic errors, we generated $N = 10,000$ trial data sets using $S = T - B = 322$ counts from an unpolarized simulation of GRB 160530A, and $B = 123$ counts from real background events taken from two 200 second intervals bracketing the GRB.  For each trial data set, the polarization analysis was performed and the resulting polarization level was stored.  Finally, the 99th percentile of polarization levels was determined numerically and found to be $\mdp = 72.3 \pm 0.8\%$.  We used this value as our 99\% confidence ($2.6 \sigma$) detection threshold.

\begin{figure}[t]
\centering
\includegraphics[width=\textwidth]{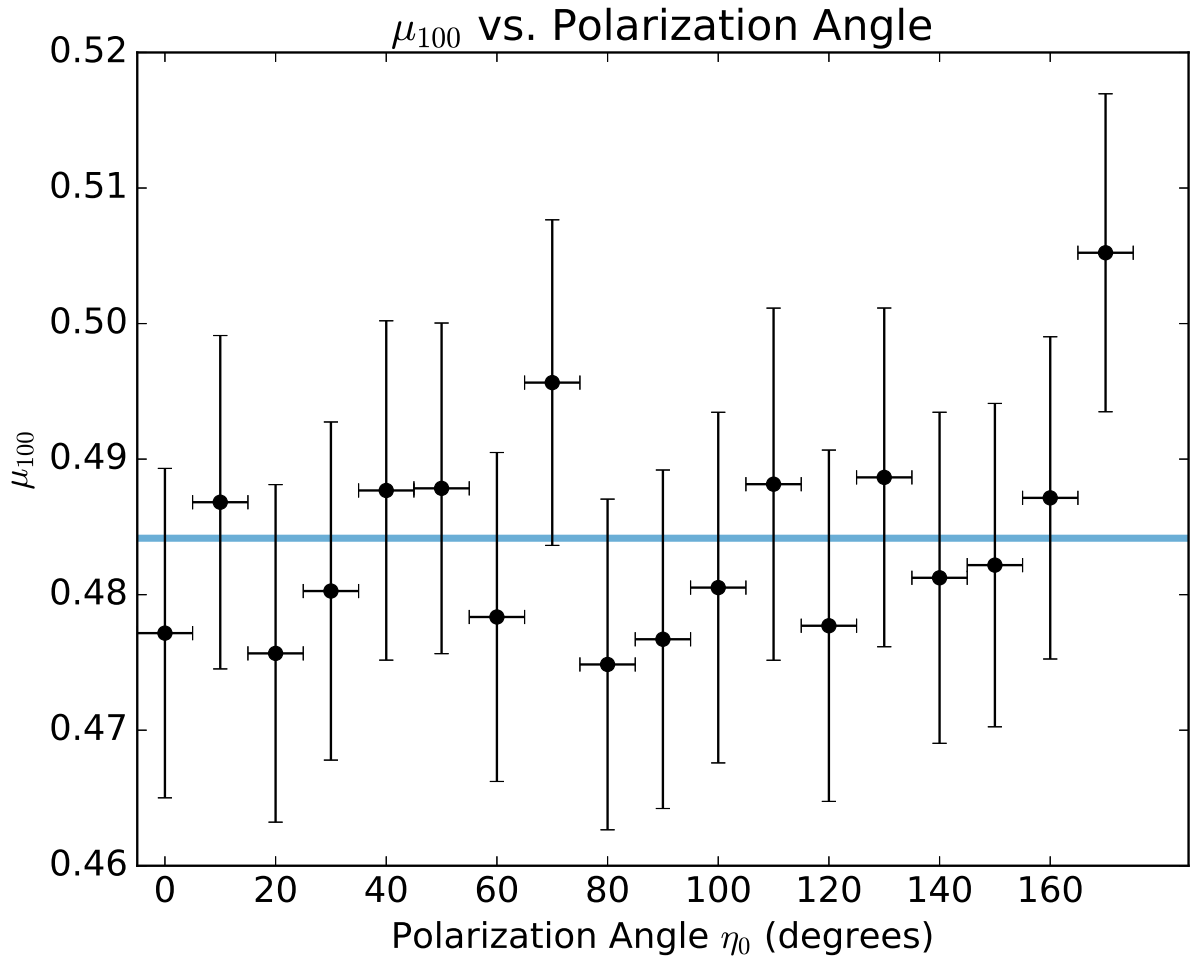}
\caption{The modulation factor $\mu_{100}$ as a function of the simulated polarization angle, along with the best fit constant value of $\mu_{100} = 0.484 \pm 0.002$, $\chi^2_{red} = 0.42$ (dof = 17).  For each simulation, $2.4-2.5 \times 10^4$ counts were used.}
\label{fig:mu100}
\end{figure}

We studied the dependence of the modulation factor $\mu_{100}$ on the polarization angle by performing MC simulations of a 100\% polarized GRB 160530A at various polarization angles.  The result of each simulation is shown in Figure \ref{fig:mu100}, along with the best fit constant.  The $\chi^2_{red}$ for the fit is 0.42 (dof = 17), indicating that the value of $\mu_{100}$ is consistent with being uniform over the full range of polarization angles.  The best fit constant  $\mu_{100} = 0.484 \pm 0.002$ was used in the following analysis.

\begin{figure}[t]
\includegraphics[width=\textwidth]{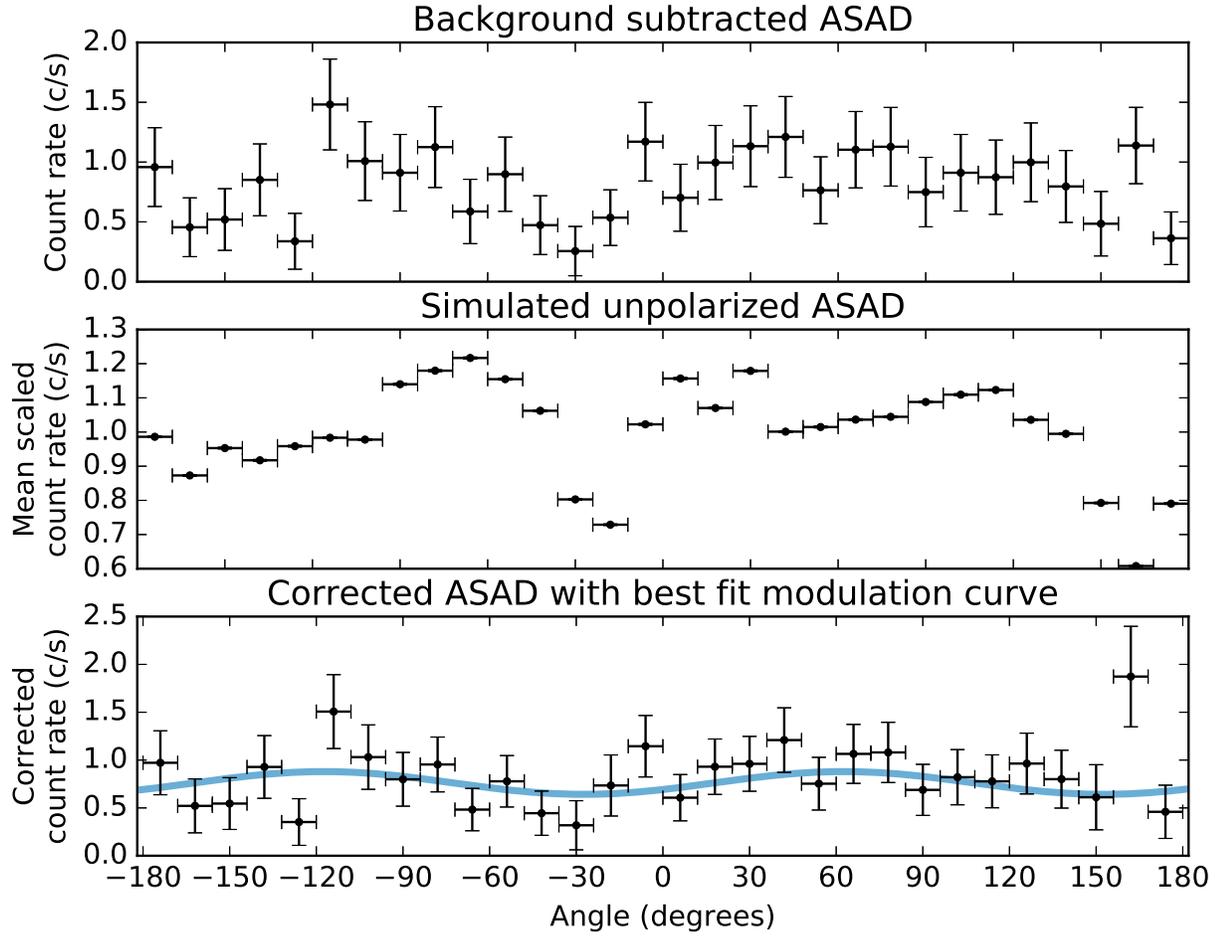}
\caption{The background subtracted azimuthal scattering angle distribution (ASAD) for GRB160530A (top), an ASAD from an unpolarized simulation of GRB160503A used to correct for systematics (middle), and the corrected ASAD (bottom) showing the best fit modulation curve in blue.}
\label{fig:asad}
\end{figure}

Once the best event selections and $\mu_{100}$ have been identified, the SM analysis may proceed.  Three separate azimuthal scattering angle distributions (ASADs) are shown in Figure \ref{fig:asad}, corresponding to a background subtracted ASAD for GRB 160530A (top), a ``correction'' ASAD generated from an unpolarized simulation of GRB 160530A (middle) which has been rescaled by the mean value, and finally the corrected ASAD with the best fit modulation curve (bottom).  The final corrected ASAD (bottom) was obtained by dividing the background-subtracted GRB 160530A ASAD (top) by the ``correction'' ASAD (middle).  This was done to correct for systematic effects of the detector system such as non-uniformity of efficiency (due to geometry, channel thresholds, etc.) and measurement uncertainty.  The fit results are shown in Table \ref{table:sm}.  We measured a polarization level of $\Pi = 33^{+33}_{-31}\%$ for this GRB using the SM, which was below the detection limit $\mdp = 72.3 \pm 0.8\%$.  In order to determine the 90\% confidence upper limit on the polarization level, we used the MINOS \citep{minos} algorithm to determine the 90\% confidence contour in the 2D parameter space of amplitude vs. offset.  Then, the maximum value for the modulation along the contour was found and divided by $\mu_{100}$, which yielded a 90\% upper limit on the polarization level of 87\%.

\subsection{Maximum Likelihood Method} \label{sec:mlm}

\begin{figure}[t]
\centering
\includegraphics[width=\textwidth]{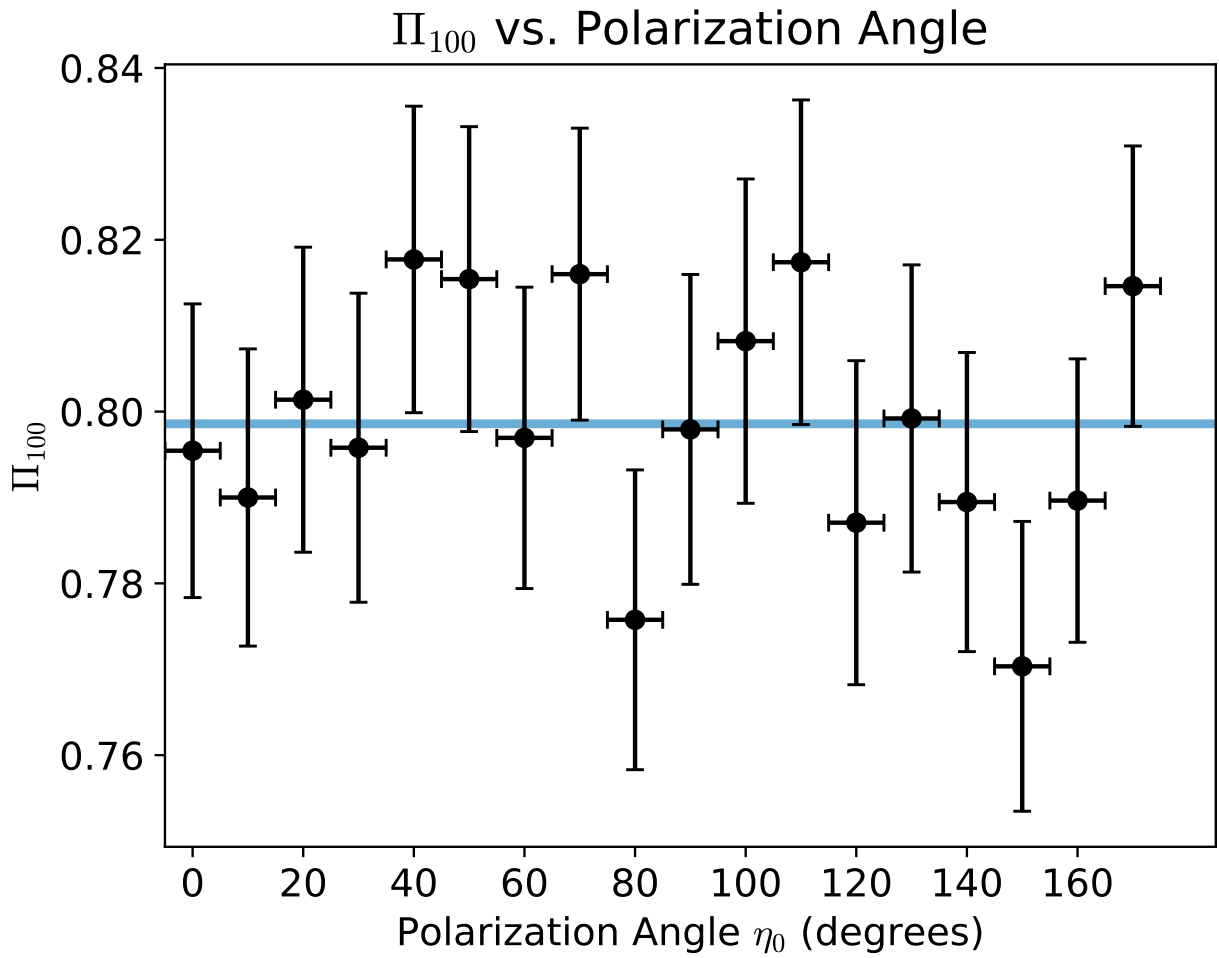}
\caption{The MLM correction factor $\Pi_{100}$ as a function of the simulated polarization angle, along with the best fit constant value $\Pi_{100} = 0.799 \pm 0.003$, $\chi^2_{red}=0.65$ (dof = 17).  For each simulation, $3.4-3.6 \times 10^4$ counts were used.}
\label{fig:pi100}
\end{figure}

The MLM aims to determine the polarization level $\Pi$ and angle $\eta_0$ that maximize the log likelihood:

\begin{equation}
\ln \Lagr = \sum_{i=1}^{N}\ln p(\eta_i;E_i,\theta_i,\Pi,\eta_0),
\label{eq:mlmsum}
\end{equation}
where $N$ is the number of events, and $p$ is the probability\footnote{\label{footnote:p}For more details on the functional form of the probability distributions used in the MLM, please refer to P2.} of measuring the azimuthal scattering angle $\eta_i$, given that the energy $E_i$ and Compton scattering angle $\theta_i$ of event $i$ have been accurately measured, and that the polarization level and angle are $\Pi$ and $\eta_0$, respectively.  The values of $\Pi$ and $\eta_0$ that maximize the log likelihood for a given observation are defined as $\hat{\Pi}$ and $\hat{\eta_0}$, where $\hat{\Pi}$ is referred to as the uncorrected polarization level.  The corrected polarization level is given by:

\begin{equation}
\Pi = \frac{\hat{\Pi}}{\Pi_{100}},
\label{eq:mlmpi}
\end{equation}
where $\Pi_{100}$ is the MLM correction factor, defined such that $\Pi_{100} = \hat{\Pi}$ when the true polarization level is $\Pi = 100\%$ and the number of background counts $B$ is zero.  An ideal polarimeter with perfect reconstruction efficiency and no measurement error would have $\Pi_{100} = 1$.  Figure \ref{fig:pi100} shows $\Pi_{100}$ for COSI as a function of polarization angle from MC simulations of GRB 160530A.  As in the SM case, the distribution is rather uniform; the best fit constant has a value of $\Pi_{100} = 0.799 \pm 0.003$ with a $\chi^2_{red}$ of 0.65 (dof = 17).  Once again, we used the best fit constant for the MLM analysis.

In the presence of background, the probability distribution in Equation \ref{eq:mlmsum} must be mixed with the probability distribution which describes the background, $p_{\mathrm{bkg}}$.  The relative contributions of the source and background probability are mixed using the signal purity $f = (T-B)/T$, where $T$ is the total number of counts, and $B$ is the expected number of background counts.  The total probability\footnote{See footnote \ref{footnote:p}.} for event $i$ is then:
\begin{equation}
p_{\textrm{total}} = f \cdot p(\eta_i;E_i,\theta_i,\Pi,\eta_0) + (1-f) \cdot p_{\mathrm{bkg}}(\eta_i;E_i,\theta_i),
\end{equation}
For the event selections in Table \ref{table:mlm}, we compute a signal purity of $f = 0.72 \pm 0.01$.  The error bar on the signal purity is the standard deviation of the simulated distribution of $f$, obtained by randomly sampling the Poisson distributions underlying $T$ and $B$.  We determined $p_{\textrm{bkg}}$ by accumulating background events from a time period which included GRB 160530A.  Due to the large number of bins (46656) in the histogram representing $p_{\textrm{bkg}}$, this time period spanned 48 hours (7 hours pre-GRB, 41 hours post-GRB).  Although it would have been preferable to only include background data from a time period closer in time to the GRB itself, this was not possible due to the relatively low absolute count rate.  The background environment at balloon float altitudes, which is dominated by cosmic ray secondaries (photons in particular), is known to vary with altitude, or equivalently, atmospheric depth \citep{ling,bowen}.  Throughout the 48 hour background integration interval, the COSI altitude was quite stable at $\sim 33.1$ km, with small excursions at a level of $<2\%$.  Thus, we conclude that using background events from this time interval is a reasonable choice.

The MLM can perform better than the SM because additional information -- the photon energy and Compton scattering angle -- is used to implicitly weight each event's contribution to the likelihood statistic \citep{kraw}.  Contrast this with the SM, where the azimuthal scattering angle of each event that passes the event selections is added to a histogram, and all other event information is discarded.  Additionally, the MLM benefits from being able to use events with any energy or Compton scattering angle, resulting in more usable counts for the analysis.  Consequently, for the MLM event selections, we accepted all photon energies between 100 and 1000 keV, as well as all Compton scattering angles $0^{\circ} - 180^{\circ}$.  With these broader selections, we determined that the FWHM of the ARM distribution was $11.9^{\circ}$ using simulated data.  For the ARM selection range, we used twice the FWHM of the ARM distribution, centered about zero, i.e. $|\textrm{ARM}| < 11.9^{\circ}$.  The distance cuts used for the MLM were the same as for the SM and for the same reasons as outlined in Section \ref{sec:sm}.

After applying the MLM event selections, $T = 542$ counts remained for the analysis, approximately $B = 152$ of which were background.  Once again, $B$ was determined by fitting a third order polynomial to the light curve, and integrating the polynomial over the time windows (Table \ref{table:times}).  The MDP was computed in a similar fashion to the SM, where $N = 10,000$ unpolarized trial observations were generated and analyzed using the MLM.  This process yielded $\mdp = 57.5 \pm 0.8\%$.

\begin{table}
\begin{center}
\begin{tabular}{ |c|c| }
\multicolumn{2}{c}{Event Selections} \\ \hline
Total photon energy & 100 - 1000 keV \\ \hline
First Compton scattering angle & $0^{\circ} - 180^{\circ}$\\ \hline
Angular resolution measure & $|\textrm{ARM}| < 11.9^{\circ}$\\ \hline
Smallest distance between first two interactions & 0.5 cm \\ \hline
Smallest distance between any two interactions & 0.3 cm \\ \hline
Number of interaction sites & 2 - 7 \\ \hline
\multicolumn{2}{c}{} \\
\multicolumn{2}{c}{Statistics} \\ \hline
Total counts $T$ & 542 \\ \hline
Background counts $B$ & 152 \\ \hline
Signal purity $f$ & $0.72 \pm 0.01$ \\ \hline
MDP & $57.5 \pm 0.8\%$ \\ \hline
MLM correction factor $\Pi_{100}$ & $0.799 \pm 0.003$ \\ \hline
\multicolumn{2}{c}{} \\
\multicolumn{2}{c}{Fit Results} \\ \hline
Fitted polarization angle $\hat{\eta_0}$ & $141^{\circ} \pm 47^{\circ}$\\ \hline
Fitted, uncorrected polarization level $\hat{\Pi}$ & $ 0.13^{+0.22}_{-0.13}$ \\ \hline
Corrected polarization level $\Pi = \frac{\hat{\Pi}}{\Pi_{100}}$ &  $16^{+27}_{-16}$ \% \\ \hline
$\Pi$ 90\% upper limit & $46\%$ \\ \hline
\end{tabular}
\end{center}
\caption{MLM event selections, statistics, and fit results.}
\label{table:mlm}
\end{table}

\begin{figure}[t]
\centering
\includegraphics[width=\textwidth]{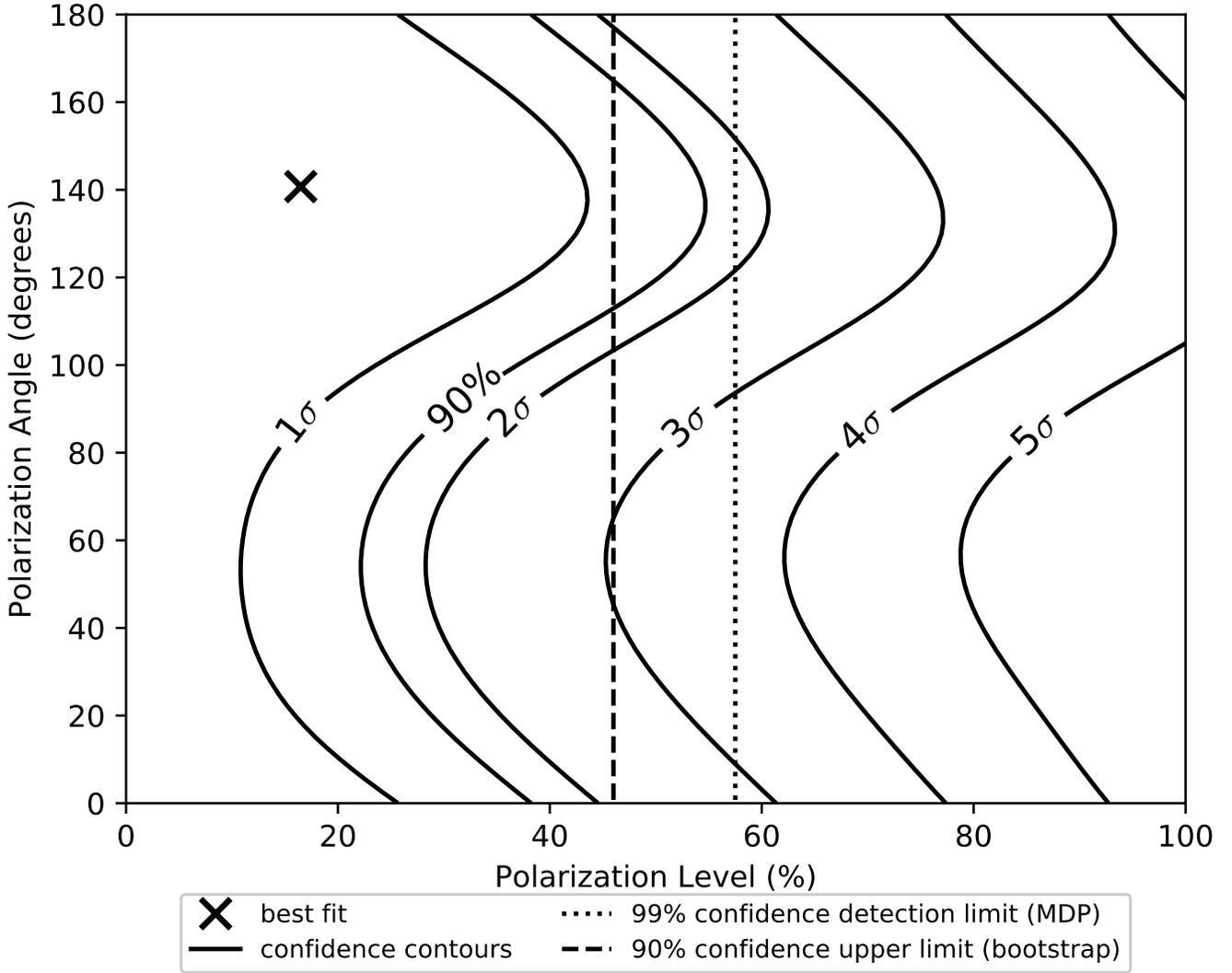}
\caption{Confidence contours for GRB 160530A, derived from a map of the log likelihood.  The optimal polarization level and angle are denoted by the cross.  Confidence contours are shown for 1-5$\sigma$ as well as 90\% confidence.  The dashed line denotes the 90\% confidence upper limit on the polarization level as determined using a Monte Carlo/bootstrap technique.  The dotted line represents the MDP, or equivalently, the 99\% confidence detection threshold.}
\label{fig:likelihood}
\end{figure}

Table \ref{table:mlm} summarizes the results of the MLM analysis, while Figure \ref{fig:likelihood} shows the confidence contours for GRB 160530A along with the best fit values, detection limit, and 90\% confidence upper limit.  The corrected polarization level was $\Pi = 16^{+27}_{-16}\%$, which was below the detection limit $\mdp = 57.5 \pm 0.8\%$.  We obtained a 90\% confidence upper limit on the corrected polarization level $\Pi$ using a Monte Carlo/bootstrap approach.  The procedure was as follows.  First we generated $N=10,000$ bootstrap resamples of the measured data.  Second, for each resample, we drew a value of $f=(T-B)/T$ by assuming a Poisson distribution for the total number of counts ($T = 542$) and the expected number of background counts ($B = 152$).  Third, we re-ran the minimizer on each resample using the value of $f$ generated in step 2 in order to yield a value of $\hat{\Pi}$.  Finally, the $\hat{\Pi}$ values were re-scaled by values of $\Pi_{100}$ drawn from its associated probability distribution, which was assumed to be Gaussian.  The 90th percentile of the resulting distribution on $\Pi$ was then used as the 90\% confidence upper limit.  Using this procedure, we arrived at a 90\% confidence upper limit of 46\% on the polarization level.  The benefit of the Monte Carlo/bootstrap approach is that the errors in the signal purity and $\Pi_{100}$ are correctly propagated into the final result.

\section{Discussion}

Both the SM and MLM give best fit polarization levels below their respective MDPs, so we conclude that COSI did not detect polarized emission from GRB 160530A at a confidence level of 99\%.  Moreover, from the error bars on the polarization levels, as well as the log likelihood contours, the measured polarization level is consistent with $\Pi=0\%$ at a confidence level of $1\sigma$.  Our most constraining 90\% confidence upper limit on the polarization level was 46\% using the MLM.  This upper limit is not strong enough to favor or disfavor any particular polarization scenario on its own.  However, several GRBs have been found to be polarized at a level above our upper limit \citep{mcconnell2017}, so our result is indeed useful in the context of the overall distribution of polarization levels within the GRB population.  GRBs with polarization levels exceeding $\Pi=46\%$ can be generated in the Compton drag model \citep{lazzati2004}, but are more frequently explained as arising from a scenario where synchrotron emission in an ordered magnetic field dominates.  However, the polarization level in such scenarios can be reduced to lower levels due to certain factors, such as an evolution in the polarization angle with time.  Such changes in the polarization angle with time were seen in GRB 100826A \citep{yonetoku2011b}, and it has been theorized that each pulse in a GRB light curve may correspond to a distinct ordering -- and thus polarization angle -- of a large scale magnetic field within the GRB jet.  Moreover, the Internal-Collision-induced MAgnetic Reconnection and Turbulence (ICMART) model for GRBs \citep{zhang2011} predicts that the polarization angle and level change from pulse to pulse, and also that the polarization level continually decreases throughout the duration of each pulse.  These changes in the polarization properties with time are theorized to result from collisions between highly magnetized shells, which are launched intermittently from the GRB central engine.  As the shells collide, the magnetic field lines distort and lose uniformity, thus reducing the polarization level of synchrotron emission.  GRB 160530A, whose early phase light curve is composed of several clear pulses, would be a good candidate to perform time-resolved polarimetry.  Unfortunately the number of counts detected with COSI from this GRB was not large enough to enable such studies.

Several factors reduced COSI's sensitivity to this GRB.  First, attenuation by the atmosphere significantly reduced the source flux; at the average measured photon energy of 360 keV, and with an effective column density of 9.2 g cm$^{-2}$, approximately 60\% of photons were absorbed or scattered by the atmosphere.  Moreover, the 100 - 1000 keV flux of the Band model with the atmospheric absorption component is reduced by 64\% compared to the unabsorbed case.  Second, the GRB occurred $43.5^{\circ}$ off axis in the COSI local frame, so the effective area was reduced by about 22\% due to occultation from the CsI shields.  Third, by the time COSI detected this GRB, two germanium detectors were no longer operational due to anomalies in the corresponding high voltage supplies.  Accordingly, all MC simulations discussed in this paper treated the two non-functioning detectors as passive mass.  A simulation of the fully functioning detector system revealed that about 16\% of events were lost due to the malfunctioning detectors.  Lastly, as mentioned in Section \ref{sec:grb}, the GRB coincided with a REP event as well as a relatively close approach to the South Magnetic Pole, both of which had the effect of elevating the background count rate.  

Despite these losses, we were able to regain some sensitivity by using the maximum likelihood based approach.  Notably, the MDP in the MLM is $\sim 20\%$ lower than the SM, while the 90\% confidence upper limit on the polarization level is $\sim 42\%$ lower.  This improvement in MDP is consistent with what was reported by \citet{kraw}, where the MDP of an idealized Compton polarimeter was shown to improve by $\sim 21\%$ when using the MLM.  

\section{Acknowledgements}

The authors would like to thank: Jamie Kennea and Phil Evans of the Swift team for their help with analysis of the Swift/XRT follow up observations; Kevin Hurley and the IPN team for their efforts to triangulate GRB160530A, which helped plan the Swift/XRT observations; Varun Bhalerao and Dipankar Bhattacharya with the Astrosat team for helping us cross-calibrate COSI's timing performance with Astrosat; the NASA Columbia Scientific Ballooning Facility team for successfully flying (and landing) COSI and making this work possible.  COSI is funded by NASA grant NNX14AC81G-APRA. Swift Guest Observer analysis is supported through NASA grant NNH15ZDA001N-SWIFT. This work is also supported in part by CNES.

\software{GEANT4 \citep{geant4}, MINUIT \citep{minos}, MEGAlib \citep{zogmegalib}, ROOT \citep{root}}

\bibliography{refs}

\begin{thebibliography}{}
\expandafter\ifx\csname natexlab\endcsname\relax\def\natexlab#1{#1}\fi
\providecommand{\url}[1]{\href{#1}{#1}}

\bibitem[{Agostinelli {et~al.}(2003)Agostinelli, Allison, Amako, Apostolakis,
  Araujo, Arce, Asai, Axen, Banerjee, Barrand, Behner, Bellagamba, Boudreau,
  Broglia, Brunengo, Burkhardt, Chauvie, Chuma, Chytracek, Cooperman, Cosmo,
  Degtyarenko, Dell'Acqua, Depaola, Dietrich, Enami, Feliciello, Ferguson,
  Fesefeldt, Folger, Foppiano, Forti, Garelli, Giani, Giannitrapani, Gibin,
  Cadenas, González, Abril, Greeniaus, Greiner, Grichine, Grossheim, Guatelli,
  Gumplinger, Hamatsu, Hashimoto, Hasui, Heikkinen, Howard, Ivanchenko,
  Johnson, Jones, Kallenbach, Kanaya, Kawabata, Kawabata, Kawaguti, Kelner,
  Kent, Kimura, Kodama, Kokoulin, Kossov, Kurashige, Lamanna, Lampén, Lara,
  Lefebure, Lei, Liendl, Lockman, Longo, Magni, Maire, Medernach, Minamimoto,
  de~Freitas, Morita, Murakami, Nagamatu, Nartallo, Nieminen, Nishimura,
  Ohtsubo, Okamura, O'Neale, Oohata, Paech, Perl, Pfeiffer, Pia, Ranjard,
  Rybin, Sadilov, Salvo, Santin, Sasaki, Savvas, Sawada, Scherer, Sei,
  Sirotenko, Smith, Starkov, Stoecker, Sulkimo, Takahata, Tanaka, Tcherniaev,
  Tehrani, Tropeano, Truscott, Uno, Urban, Urban, Verderi, Walkden, Wander,
  Weber, Wellisch, Wenaus, Williams, Wright, Yamada, Yoshida, \&
  Zschiesche}]{geant4}
Agostinelli, S., Allison, J., Amako, K., {et~al.} 2003, Nuclear Instruments and
  Methods in Physics Research Section A: Accelerators, Spectrometers, Detectors
  and Associated Equipment, 506, 250 .
\newblock
  \url{http://www.sciencedirect.com/science/article/pii/S0168900203013688}

\bibitem[{Amman {et~al.}(2007)Amman, Luke, \& Boggs}]{amman2007}
Amman, M., Luke, P., \& Boggs, S. 2007, Nuclear Instruments and Methods in
  Physics Research Section A: Accelerators, Spectrometers, Detectors and
  Associated Equipment, 579, 886 .
\newblock
  \url{http://www.sciencedirect.com/science/article/pii/S0168900207012004}

\bibitem[{{Band} {et~al.}(1993){Band}, {Matteson}, {Ford}, {Schaefer},
  {Palmer}, {Teegarden}, {Cline}, {Briggs}, {Paciesas}, {Pendleton}, {Fishman},
  {Kouveliotou}, {Meegan}, {Wilson}, \& {Lestrade}}]{band}
{Band}, D., {Matteson}, J., {Ford}, L., {et~al.} 1993, \apj, 413, 281

\bibitem[{Bandstra {et~al.}(2009)Bandstra, Bellm, Chiu, Liang, Liu,
  Perez-Becker, Zoglauer, Boggs, Chang, Chang, {et~al.}}]{bandstra2009spring}
Bandstra, M.~S., Bellm, E.~C., Chiu, J.-L., {et~al.} 2009, in Nuclear Science
  Symposium Conference Record (NSS/MIC), 2009 IEEE, IEEE, 2131--2139

\bibitem[{{Bandstra} {et~al.}(2011){Bandstra}, {Bellm}, {Boggs},
  {Perez-Becker}, {Zoglauer}, {Chang}, {Chiu}, {Liang}, {Chang}, {Liu}, {Hung},
  {Huang}, {Chiang}, {Run}, {Lin}, {Amman}, {Luke}, {Jean}, {von Ballmoos}, \&
  {Wunderer}}]{bandstracrab}
{Bandstra}, M.~S., {Bellm}, E.~C., {Boggs}, S.~E., {et~al.} 2011, \apj, 738, 8

\bibitem[{Bhalerao {et~al.}(2017)Bhalerao, Bhattacharya, Vibhute, Pawar, Rao,
  Hingar, Khanna, Kutty, Malkar, Patil, Arora, Sinha, Priya, Samuel, Sreekumar,
  Vinod, Mithun, Vadawale, Vagshette, Navalgund, Sarma, Pandiyan, Seetha, \&
  Subbarao}]{czti}
Bhalerao, V., Bhattacharya, D., Vibhute, A., {et~al.} 2017, Journal of
  Astrophysics and Astronomy, 38, 31.
\newblock \url{https://doi.org/10.1007/s12036-017-9447-8}

\bibitem[{Boggs {et~al.}(2006)Boggs, Bandstra, Bowen, Coburn, Lin, Wunderer,
  Zoglauer, Amman, Luke, Jean, {et~al.}}]{boggs2006}
Boggs, S., Bandstra, M., Bowen, J., {et~al.} 2006, in Focusing Telescopes in
  Nuclear Astrophysics (Springer), 387--394

\bibitem[{Bowen {et~al.}(2007)Bowen, Bandstra, Boggs, Zoglauer, Wunderer,
  Amman, \& Luke}]{bowen}
Bowen, J.~D., Bandstra, M.~E., Boggs, S.~E., {et~al.} 2007, in 2007 IEEE
  Nuclear Science Symposium Conference Record, Vol.~1, 436--444

\bibitem[{Brun \& Rademakers(1997)}]{root}
Brun, R., \& Rademakers, F. 1997, Nuclear Instruments and Methods in Physics
  Research Section A: Accelerators, Spectrometers, Detectors and Associated
  Equipment, 389, 81 , new Computing Techniques in Physics Research V.
\newblock
  \url{http://www.sciencedirect.com/science/article/pii/S016890029700048X}

\bibitem[{{Chattopadhyay} {et~al.}(2017){Chattopadhyay}, {Vadawale}, {Aarthy},
  {Mithun}, {Chand}, {Basak}, {Rao}, {Mate}, {Sharma}, {Bhalerao}, \&
  {Bhattacharya}}]{astrosatgrb}
{Chattopadhyay}, T., {Vadawale}, S.~V., {Aarthy}, E., {et~al.} 2017, ArXiv
  e-prints, arXiv:1707.06595

\bibitem[{Chiu {et~al.}(2015)Chiu, Boggs, Chang, Tomsick, Zoglauer, Amman,
  Chang, Chou, Jean, Kierans, Lin, Lowell, Shang, Tseng, von Ballmoos, \&
  Yang}]{chiu2015}
Chiu, J.-L., Boggs, S., Chang, H.-K., {et~al.} 2015, Nuclear Instruments and
  Methods in Physics Research Section A: Accelerators, Spectrometers, Detectors
  and Associated Equipment, 784, 359 , symposium on Radiation Measurements and
  Applications 2014 (SORMA XV).
\newblock
  \url{http://www.sciencedirect.com/science/article/pii/S0168900214014132}

\bibitem[{{D'Elia} {et~al.}(2016{\natexlab{a}}){D'Elia}, {Izzo}, {Breeveld}, \&
  {Markwardt}}]{swift3gcn}
{D'Elia}, V., {Izzo}, L., {Breeveld}, A.~A., \& {Markwardt}, C.~B.
  2016{\natexlab{a}}, GRB Coordinates Network, 19501

\bibitem[{{D'Elia} {et~al.}(2016{\natexlab{b}}){D'Elia}, {Izzo}, {Evans},
  {Breeveld}, {Kennea}, {Malesani}, \& {Markwardt}}]{swift2gcn}
{D'Elia}, V., {Izzo}, L., {Evans}, P.~A., {et~al.} 2016{\natexlab{b}}, GRB
  Coordinates Network, 19481

\bibitem[{{Evans}(2016)}]{swift1gcn}
{Evans}, P.~A. 2016, GRB Coordinates Network, 19472

\bibitem[{Granot(2003)}]{granot2003}
Granot, J. 2003, Astrophysical Journal, 596, L17.
\newblock \url{www.scopus.com}

\bibitem[{{Gruzinov} \& {Waxman}(1999)}]{gruzinov1999}
{Gruzinov}, A., \& {Waxman}, E. 1999, \apj, 511, 852

\bibitem[{James \& Roos(1975)}]{minos}
James, F., \& Roos, M. 1975, Computer Physics Communications, 10, 343

\bibitem[{{Kierans} {et~al.}(2017){Kierans}, {Boggs}, {Chiu}, {Lowell},
  {Sleator}, {Tomsick}, {Zoglauer}, {Amman}, {Chang}, {Tseng}, {Yang}, {Lin},
  {Jean}, \& {von Ballmoos}}]{kierans2017}
{Kierans}, C.~A., {Boggs}, S.~E., {Chiu}, J.-L., {et~al.} 2017, ArXiv e-prints,
  arXiv:1701.05558

\bibitem[{{Kole} \& {Marcinkowski}(2016)}]{polargcn1}
{Kole}, M., \& {Marcinkowski}, R. 2016, GRB Coordinates Network, 20301

\bibitem[{{Kole} {et~al.}(2016){Kole}, {Produit}, {Bernasconi}, {Li}, {Zhao},
  {Batsch}, {Cadoux}, {Dong}, {Feng}, {Ge}, {Hajdas}, {Huang}, {Jedrzejczak},
  {Lu}, {Marcinkowski}, {Pohl}, {Rutczynska}, {Rybka}, {Song}, {Szabelski},
  {Wen}, {Wu}, {Wu}, {Xiao}, {Xu}, {Zhang}, {Zhang}, {Zhang}, {Zhang}, {Zhang},
  \& {Zwolinska}}]{polargcn2}
{Kole}, M., {Produit}, N., {Bernasconi}, T., {et~al.} 2016, GRB Coordinates
  Network, 20243

\bibitem[{Krawczynski(2011)}]{kraw}
Krawczynski, H. 2011, Astroparticle Physics, 34, 784

\bibitem[{Lazzati {et~al.}(2004)Lazzati, Rossi, Ghisellini, \&
  Rees}]{lazzati2004}
Lazzati, D., Rossi, E., Ghisellini, G., \& Rees, M.~J. 2004, Monthly Notices of
  the Royal Astronomical Society, 347, L1.
\newblock \url{+ http://dx.doi.org/10.1111/j.1365-2966.2004.07387.x}

\bibitem[{Lei {et~al.}(1997)Lei, Dean, \& Hills}]{lei1997}
Lei, F., Dean, A., \& Hills, G. 1997, Space Science Reviews, 82, 309

\bibitem[{{Ling}(1975)}]{ling}
{Ling}, J.~C. 1975, \jgr, 80, 3241

\bibitem[{Lowell {et~al.}(2016)Lowell, Boggs, Chiu, Kierans, McBride, Tseng,
  Zoglauer, Amman, Chang, Jean, Lin, Sleator, Tomsick, von Ballmoos, \&
  Yang}]{lowell2016}
Lowell, A., Boggs, S., Chiu, J.~L., {et~al.} 2016, Positional calibrations of
  the germanium double sided strip detectors for the Compton spectrometer and
  imager, , , doi:10.1117/12.2233145.
\newblock \url{http://dx.doi.org/10.1117/12.2233145}

\bibitem[{Lyutikov {et~al.}(2003)Lyutikov, Pariev, \& Blandford}]{lyutikov2003}
Lyutikov, M., Pariev, V.~I., \& Blandford, R.~D. 2003, Astrophysical Journal,
  597, 998.
\newblock \url{www.scopus.com}

\bibitem[{McConnell(2017)}]{mcconnell2017}
McConnell, M.~L. 2017, New Astronomy Reviews, 76, 1 .
\newblock
  \url{http://www.sciencedirect.com/science/article/pii/S1387647316300495}

\bibitem[{Orsi {et~al.}(2014)Orsi, Cadoux, Leluc, Paniccia, Pohl, Rapin,
  Gauvin, Produit, Bao, Chai, Dong, Kong, Lu, Liu, Liu, Shi, Sun, Wang, Wen,
  Wu, Xiao, Xu, Zhang, Zhang, Zhang, Zhang, Britvich, Hajdas, Marcinkowski,
  Rybka, Batsch, Rutczynska, Szabelski, \& Zwolinska}]{orsi2014}
Orsi, S., Cadoux, F., Leluc, C., {et~al.} 2014, The POLAR gamma-ray burst
  polarimeter onboard the Chinese Spacelab, , , doi:10.1117/12.2055910.
\newblock \url{http://dx.doi.org/10.1117/12.2055910}

\bibitem[{Parks {et~al.}(1979)Parks, Gurgiolo, \& West}]{parksrep}
Parks, G., Gurgiolo, C., \& West, R. 1979, Geophysical Research Letters, 6, 393

\bibitem[{Picone {et~al.}(2002)Picone, Hedin, Drob, \& Aikin}]{nrlmsise}
Picone, J., Hedin, A., Drob, D.~P., \& Aikin, A. 2002, Journal of Geophysical
  Research: Space Physics, 107

\bibitem[{Produit {et~al.}(2005)Produit, Barao, Deluit, Hajdas, Leluc, Pohl,
  Rapin, Vialle, Walter, \& Wigger}]{produit2005}
Produit, N., Barao, F., Deluit, S., {et~al.} 2005, Nuclear Instruments and
  Methods in Physics Research Section A: Accelerators, Spectrometers, Detectors
  and Associated Equipment, 550, 616

\bibitem[{{Shaviv} \& {Dar}(1995)}]{shaviv1995}
{Shaviv}, N.~J., \& {Dar}, A. 1995, \apj, 447, 863

\bibitem[{Sleator {et~al.}(2017)Sleator, Boggs, Chiu, Kierans, Lowell, Tomsick,
  Zoglauer, Amman, Chang, Tseng, {et~al.}}]{sleator2017}
Sleator, C.~C., Boggs, S.~E., Chiu, J.-L., {et~al.} 2017, arXiv preprint
  arXiv:1701.05563

\bibitem[{Storn \& Price(1997)}]{Storn1997}
Storn, R., \& Price, K. 1997, Journal of Global Optimization, 11, 341.
\newblock \url{http://dx.doi.org/10.1023/A:1008202821328}

\bibitem[{Strohmayer \& Kallman(2013)}]{strohmayer}
Strohmayer, T.~E., \& Kallman, T.~R. 2013, The Astrophysical Journal, 773, 103.
\newblock \url{http://stacks.iop.org/0004-637X/773/i=2/a=103}

\bibitem[{Sun {et~al.}(2016)Sun, Wu, Bao, Batsch, Bernasconi, Britvitch,
  Cadoux, Cernuda, Chai, Dong, Gauvin, Hajdas, He, Kole, Kong, Kong,
  Lechanoine-Leluc, Li, Liu, Liu, Marcinkowski, Orsi, Pohl, Produit, Rapin,
  Rutczynska, Rybka, Shi, Song, Szabelski, Wang, Wen, Xiao, Xiong, Xu, Xu,
  Zhang, Zhang, Zhang, Zhang, Zhang, \& Zwolinska}]{sun2016}
Sun, J.~C., Wu, B.~B., Bao, T.~W., {et~al.} 2016, Performance study of the
  gamma-ray bursts polarimeter POLAR, , , doi:10.1117/12.2232133.
\newblock \url{http://dx.doi.org/10.1117/12.2232133}

\bibitem[{{Svinkin} {et~al.}(2016{\natexlab{a}}){Svinkin}, {Golenetskii},
  {Aptekar}, {Frederiks}, {Kozlova}, {Cline}, {Hurley}, {von Kienlin}, {Zhang},
  {Rau}, {Savchenko}, {Bozzo}, \& {Ferrigno}}]{triangulationgcn}
{Svinkin}, D., {Golenetskii}, S., {Aptekar}, R., {et~al.} 2016{\natexlab{a}},
  GRB Coordinates Network, 19476

\bibitem[{{Svinkin} {et~al.}(2016{\natexlab{b}}){Svinkin}, {Golenetskii},
  {Aptekar}, {Frederiks}, {Oleynik}, {Ulanov}, {Tsvetkova}, {Lysenko},
  {Kozlova}, \& {Cline}}]{spectrumgcn}
---. 2016{\natexlab{b}}, GRB Coordinates Network, 19477

\bibitem[{Toma {et~al.}(2009)Toma, Sakamoto, Zhang, Hill, McConnell, Bloser,
  Yamazaki, Ioka, \& Nakamura}]{toma2009}
Toma, K., Sakamoto, T., Zhang, B., {et~al.} 2009, The Astrophysical Journal,
  698, 1042

\bibitem[{Tomsick \& the COSI~team(2016)}]{cosigcn}
Tomsick, J.~A., \& the COSI~team. 2016, GRB Coordinates Network, 19473

\bibitem[{{Waxman}(2003)}]{waxman2003}
{Waxman}, E. 2003, \nat, 423, 388

\bibitem[{{Weisskopf} {et~al.}(2010){Weisskopf}, {Elsner}, \&
  {O'Dell}}]{weisskopf}
{Weisskopf}, M.~C., {Elsner}, R.~F., \& {O'Dell}, S.~L. 2010, in \procspie,
  Vol. 7732, Space Telescopes and Instrumentation 2010: Ultraviolet to Gamma
  Ray, 77320E

\bibitem[{Wilderman {et~al.}(1998)Wilderman, Clinthorne, Fessler, \&
  Rogers}]{wilderman1998}
Wilderman, S.~J., Clinthorne, N.~H., Fessler, J.~A., \& Rogers, W.~L. 1998, in
  Nuclear Science Symposium, 1998. Conference Record. 1998 IEEE, Vol.~3, IEEE,
  1716--1720

\bibitem[{Yonetoku {et~al.}(2011{\natexlab{a}})Yonetoku, Murakami, Gunji,
  Mihara, Sakashita, Morihara, Kikuchi, Takahashi, Fujimoto, Toukairin, Kodama,
  Kubo, \& }]{yonetoku2011}
Yonetoku, D., Murakami, T., Gunji, S., {et~al.} 2011{\natexlab{a}},
  Publications of the Astronomical Society of Japan, 63, 625.
\newblock \url{+ http://dx.doi.org/10.1093/pasj/63.3.625}

\bibitem[{Yonetoku {et~al.}(2011{\natexlab{b}})Yonetoku, Murakami, Gunji,
  Mihara, Toma, Sakashita, Morihara, Takahashi, Toukairin, Fujimoto, Kodama,
  Kubo, \& Team}]{yonetoku2011b}
---. 2011{\natexlab{b}}, The Astrophysical Journal Letters, 743, L30.
\newblock \url{http://stacks.iop.org/2041-8205/743/i=2/a=L30}

\bibitem[{Yonetoku {et~al.}(2012)Yonetoku, Murakami, Gunji, Mihara, Toma,
  Morihara, Takahashi, Wakashima, Yonemochi, Sakashita, Toukairin, Fujimoto, \&
  Kodama}]{yonetoku2012}
---. 2012, The Astrophysical Journal Letters, 758, L1.
\newblock \url{http://stacks.iop.org/2041-8205/758/i=1/a=L1}

\bibitem[{Zhang \& Yan(2011)}]{zhang2011}
Zhang, B., \& Yan, H. 2011, The Astrophysical Journal, 726, 90.
\newblock \url{http://stacks.iop.org/0004-637X/726/i=2/a=90}

\bibitem[{{Zoglauer} {et~al.}(2006){Zoglauer}, {Andritschke}, \&
  {Schopper}}]{zogmegalib}
{Zoglauer}, A., {Andritschke}, R., \& {Schopper}, F. 2006, \nar, 50, 629

\bibitem[{Zoglauer(2005)}]{zoglauerthesis}
Zoglauer, A.~C. 2005, PhD thesis, Technische Universit{\"a}t M{\"u}nchen

\end{thebibliography}

\end{document}